\documentclass[aps,prd,twocolumn,superscriptaddress,nofootinbib]{revtex4-1}

\usepackage{latexsym}
\usepackage{amsmath}
\usepackage{amssymb}
\usepackage{mathtools,slashed}
\usepackage{amsfonts}
\usepackage{bm}
\usepackage{physics}
\usepackage[table,x11names]{xcolor}
\usepackage{color}
\definecolor{purple}{rgb}{0.5,0,0.5}
\definecolor{blue}{rgb}{0.0,0,0.9}
\definecolor{prdblue}{rgb}{0.133,0.118,0.498}
\usepackage[colorlinks=true, pdfstartview=FitV, linkcolor=prdblue, citecolor= prdblue, urlcolor=prdblue]{hyperref}

\usepackage{supertabular} 
\usepackage{placeins}
\usepackage{epsfig}
\usepackage{graphicx}

\def\tstrut{\vrule height3.25ex depth0pt width0pt} 

\begin{document}


\title{Pseudoscalar and vector toponia in a Dyson--Schwinger--Bethe--Salpeter framework}


\author{H.-R. Zhang}
\email[]{hrzhang@smail.nju.edu.cn}
\affiliation{School of Physics, Nanjing University, Nanjing, Jiangsu 210093, China}

\author{Z.-F. Cui}
\email[]{phycui@nju.edu.cn}
\affiliation{School of Physics, Nanjing University, Nanjing, Jiangsu 210093, China}

\author{J. Segovia}
\email[]{jsegovia@upo.es}
\affiliation{Dpto. Sistemas F\'isicos, Qu\'imicos y Naturales, Univ. Pablo de Olavide, 41013 Sevilla, Spain}

\date{\today}

\begin{abstract}
We study the pseudoscalar ($J^{PC}=0^{-+}$) and vector ($1^{--}$) top--antitop (toponium) systems within the rainbow--ladder truncation of the Dyson--Schwinger and Bethe--Salpeter equations, employing the Qin--Chang effective interaction. After validating the framework in the charmonium and bottomonium sectors, we extend it consistently to the top sector, incorporating renormalisation-group running of the current quark mass and a careful treatment of the number of active flavours. We compute masses and leptonic decay constants for $N_f=5$ and $6$, then analyse their dependence on the renormalisation scale in the range $\mu=400-800$~GeV. The resulting toponium masses lie near $344-346\,\text{GeV}$ with hyperfine splittings below $0.14-0.17\,\text{GeV}$, while the decay constants are large, $6-7\,\text{GeV}$, and exhibit the expected heavy-quark scaling behaviour. We find only mild sensitivity to the renormalisation point and a systematic reduction of binding when increasing $N_f$. Although the physical top quark decays weakly before hadronisation, our results demonstrate that, within a Poincar\'e-covariant nonperturbative framework, quantum chromodynamics (QCD) generates tightly correlated pseudoscalar and vector toponium systems in that extreme heavy-quark limit.
\end{abstract}


\maketitle


\section{INTRODUCTION}
\label{sec:introduction}

The top quark, $t$, with a mass of about $173\,\text{GeV}$, is the heaviest known elementary particle and represents a unique laboratory for testing the Standard Model at the interface between electroweak and strong interactions~\cite{ParticleDataGroup:2024cfk}. Its large mass leads to an extraordinarily short lifetime, $\tau_t \sim 10^{-25}\,\text{s} < \Lambda_{\text{QCD}}^{-1}\sim 10^{-24}\,\text{s}$, such that the top-quark typically decays via the weak interaction before conventional hadronisation can occur. As a consequence, top--anti-top ($t\bar t$, toponium) bound states analogous to charmonium ($c\bar c$) and bottomonium ($b\bar b$) have long been considered inaccessible via experiment~\cite{Aguilar-Saavedra:2024mnm}. 

Nevertheless, recent analyses by the CMS Collaboration of proton--proton collision data at $\sqrt{s}=13\,\text{TeV}$ have revealed a significant excess of $t\bar t$ events near the kinematic threshold that is compatible with the production of a color-singlet pseudoscalar quasi-bound state, as suggested by non-relativistic QCD (NRQCD) calculations for toponium, with a measured cross section of approximately $8.8^{+1.2}_{-1.4}\,\text{pb}$ for the excess above the fixed-order prediction of perturbative QCD at the threshold~\cite{CMS:2025kzt}. Subsequent reports from the ATLAS Collaboration confirm an enhancement near threshold, with a measured cross section consistent with the CMS observation and a combined significance exceeding that of background-only models~\cite{ATLAS:2025mvr}.

These unforeseen features in the top-pair production data have rejuvenated interest in the theoretical description of heavy quark--antiquark correlations in the extreme mass regime of the top quark~\cite{Wang:2024hzd, Llanes-Estrada:2024phk, Fu:2025yft, Najjar:2025bby, Zhang:2026mmc}. In contrast to the conventional expectation that the top decay width $\Gamma_t$ is too large for formation of narrow hadronic bound states, the interplay between QCD binding effects and electroweak instability calls for a careful treatment of near-threshold dynamics~\cite{Fuks:2024yjj, Fuks:2025toq}. Theoretical approaches based on NRQCD have provided a useful phenomenological description of near-threshold behaviour~\cite{Beneke:2013jia, Beneke:2024sfa, Garzelli:2024uhe, Shao:2025dzw}, but a fully relativistic and Poincar\'e-covariant treatment is still needed to study toponium in a first-principles framework.

In the continuum formulation of QCD, meson properties can be studied through the Dyson--Schwinger equations (DSE) for the dressed quark and gluon propagators together with the homogeneous Bethe--Salpeter equation (BSE) for quark--antiquark bound states~\cite{Roberts:1994dr, Alkofer:2000wg, Maris:2003vk, Qin:2020rad}. These equations form an infinite tower of coupled integral equations derived from the QCD generating functional~\cite{Huber:2011qr, Huber:2018ned}. When truncated in a symmetry-preserving manner~\cite{Eichmann:2008ae, Binosi:2016rxz, Qin:2019oar}, they provide a Poincar\'e-covariant and renormalisation-group consistent framework for describing mesons, and baryons, across all quark-mass regimes~\cite{Qin:2011xq, Chang:2013nia, Raya:2015gva, Eichmann:2016yit, Lu:2017cln, Chen:2019fzn, Lu:2019bjs, Yin:2019bxe, Yin:2021uom, Liu:2022ndb}.

A particularly successful construction of the effective quark--antiquark interaction -- often referred to as the Qin--Chang interaction -- was developed to provide a unified description of meson observables across a wide range of quark masses~\cite{Qin:2011dd}. In this model the interaction kernel combines an infrared-enhanced term responsible for dynamical chiral symmetry breaking with a perturbative ultraviolet tail consistent with QCD's one-loop behaviour. When implemented in the rainbow--ladder (RL) truncation of the quark gap and meson's BSEs, this interaction systematically reproduces meson pseudoscalar and vector masses and leptonic decay constants in the light~\cite{Qin:2020jig}, heavy-light~\cite{Xu:2022kng}, and heavy sectors, including charmonium and bottomonium~\cite{Qin:2018dqp}. Moreover, such studies preserve the axial-vector Ward--Green--Takahashi identity~\cite{Qin:2014vya, Qin:2013mta}, ensuring the correct interplay between chiral symmetry and bound-state structure.

In this work, we explore the lowest-lying pseudoscalar $(J^{PC}=0^{-+})$ and vector $(1^{--})$ toponium systems within the DSE-BSE framework using the Qin--Chang effective interaction. Building on earlier successes in the description of heavy quarkonia, we extend the model consistently to the top sector by including renormalisation-group running of the current quark mass, careful treatment of the number of active flavours and systematic analysis of renormalisation-scale dependence. We compute the masses and leptonic decay constants for $N_f=5$ and $6$ active flavours and analyse their behaviour in a range of renormalisation points relevant for heavy quark dynamics. These results provide a fully relativistic baseline for the mass spectrum and decay constants of hypothetical toponium states, and offer insight into how QCD dynamics manifests itself when the quark lifetime competes with binding effects. In the present study we do not include explicit finite-width effects associated with the electroweak decay of the top quark. Our results therefore characterise the strength of QCD-induced quark--antiquark correlations in the extreme heavy-quark limit, providing a baseline against which more refined treatments incorporating unstable-particle dynamics can be compared.

The manuscript is organised as follows. In Sec.~\ref{sec:theory} we introduce the Dyson--Schwinger--Bethe--Salpeter formalism and specify the Qin--Chang interaction employed in this work. Section~\ref{sec:results} presents our results, first discussing the treatment of the running top-quark mass, second validating the framework in the charmonium and bottomonium sectors and third presenting our results for pseudoscalar and vector toponium states for $N_f=5$ and $6$, including the analysis of renormalisation-scale dependence and heavy-quark scaling. Finally, Sec.~\ref{sec:summary} summarises our findings and outlines possible extensions, such as the completion of toponium spectrum, inclusion of finite-width effects and beyond-rainbow corrections.


\section{Theoretical Formalism}
\label{sec:theory}

The RL truncation, which replaces the fully dressed quark--gluon vertex with an effective interaction while maintaining consistency between the quark DSE and meson BSE kernels, preserves the axial-vector Ward--Green--Takahashi identity. As a consequence, chiral symmetry and its dynamical breaking are correctly realised, ensuring that pseudoscalar mesons emerge as (pseudo-)Goldstone bosons in the light-quark sector. At the same time, RL truncation provides a quantitatively reliable description of heavy quarkonia, where corrections beyond ladder exchange are suppressed by inverse powers of the heavy-quark mass.

A widely used realisation of this framework is the effective interaction introduced by Qin and Chang et al~\cite{Qin:2011dd}. In this construction, the interaction kernel combines an infrared-enhanced term responsible for dynamical chiral symmetry breaking with an ultraviolet component constrained to reproduce the one-loop renormalisation-group behaviour of QCD. The model is characterised by two parameters, $\omega$ and $D$, which control the infrared width and strength of the interaction and have been shown to yield a unified description of pseudoscalar and vector mesons from the light sector through charmonium and bottomonium. In particular, masses and leptonic decay constants of $\eta_c$, $J/\psi$, $\eta_b$, and $\Upsilon$ are reproduced at the few-percent level within a common parameter set~\cite{Qin:2018dqp}. Extending the framework to the top quark therefore probes the extreme heavy-mass limit of QCD within a fully relativistic, Poincar\'e-covariant formalism.

\subsection{Quark Dyson--Schwinger Equation}

In continuum QCD, the dressed quark propagator is obtained from the quark DSE,\footnote{Note that we employ Euclidean conventions throughout this study with $\{\gamma_\mu,\gamma_\nu\} = 2\delta_{\mu\nu}$, $\gamma^\dagger_\mu = \gamma_\mu$ and
$a \cdot b =\sum_{i=1}^4 a_i b_i$ and a spacelike vector, $k_\mu$, has $k^2>0$.}
\begin{align}
S^{-1}(p) &= Z_2 \, ( i \slashed{p} + m^{\rm bm} ) \nonumber \\
&
+ Z_1 \, g^2 \int^\Lambda_k D_{\mu\nu}(q) \, \frac{\lambda^a}{2} \gamma_\mu \,
S(k) \, \Gamma^a_\nu(k,p) \,,
\label{eq:quarkDSE}
\end{align}
where $Z_1$, $Z_2$ are renormalisation constants, $m^{\rm bm}$ is the bare mass, $D_{\mu\nu}$ is the dressed gluon propagator, $q=p-k$, and $\Gamma^a_\nu$ is the dressed quark--gluon vertex. The integral symbol denotes
\begin{equation}
\int^\Lambda_k = \int^\Lambda \frac{d^4k}{(2\pi)^4} \,,
\end{equation}
regularised with ultraviolet cutoff $\Lambda$.

The renormalised quark propagator can be decomposed as
\begin{equation}
S(p)^{-1} = i \slashed{p} A(p^2,\mu^2) + B(p^2,\mu^2) \,,
\end{equation}
or equivalently,
\begin{equation}
S(p) = \frac{Z(p^2,\mu^2)} {i \slashed{p} + M(p^2)},
\end{equation}
where $M(p^2)=B/A$ is the running mass function and $\mu$ is the renormalisation point.

Renormalisation is implemented via
\begin{equation}
S^{-1}(p)\big|_{p^2=\mu^2} = i \slashed{p} + m(\mu) \,,
\end{equation}
ensuring consistency with the chosen renormalisation scheme.

\subsection{Rainbow--Ladder Truncation}

In the RL truncation, the dressed quark--gluon vertex is replaced by its leading Dirac structure and an effective interaction,
\begin{equation}
Z_1 g^2 D_{\mu\nu}(q)\Gamma^a_\nu(k,p) \rightarrow 4\pi \alpha_{\rm eff}(q^2)
D^{\rm free}_{\mu\nu}(q) \frac{\lambda^a}{2} \gamma_\nu \,,
\end{equation}
where $D^{\rm free}_{\mu\nu}(q)$ is the free gluon propagator in Landau-gauge,
\begin{equation}
D^{\rm free}_{\mu\nu}(q) = \frac{1}{q^2} \left( \delta_{\mu\nu} - \frac{q_\mu q_\nu}{q^2} \right) \,.
\end{equation}

This truncation preserves the axial-vector Ward--Green--Takahashi identity,
\begin{equation}
P_\mu \Gamma_{5\mu}(k;P) = S^{-1}(k_+)\gamma_5 + \gamma_5 S^{-1}(k_-) \,,
\end{equation}
where $k_\pm=k\pm P/2$, with $k$ the relative momentum and $P$ the total momentum. This ensures the correct realisation of chiral symmetry and its dynamical breaking.

\subsection{Qin--Chang Effective Interaction}

The effective coupling $\alpha_{\rm eff}(q^2)$ is taken in the Qin--Chang form,
\begin{align}
\frac{\alpha_{\rm eff}(q^2)}{q^2} &= \frac{2\pi D}{\omega^4} e^{-q^2/\omega^2} \nonumber \\
&
+ \frac{2\pi \gamma_m \left(1 - e^{-q^2/\Lambda_t^2}\right)} {q^2 \ln\!\left[ e^2 -1 + \left(1+ q^2/\Lambda^2_{\rm QCD}\right)^2 \right]} \,,
\label{eq:QCinteraction}
\end{align}
where $D$ and $\omega$ control the infrared strength and width, $\gamma_m = 12/(33-2N_f)$ is the anomalous dimension, with $N_f$ the number of active flavors, $\Lambda_{\rm QCD}$ sets the ultraviolet scale, $\Lambda_t=1.0\,\text{GeV}$ ensures smooth infrared matching.

The first term in Eq.~\eqref{eq:QCinteraction} generates dynamical chiral symmetry breaking, while the second reproduces the one-loop QCD running at large momenta.

\subsection{Bethe--Salpeter Equation}

Mesons appear as poles in the quark--antiquark scattering matrix. The homogeneous BSE reads
\begin{equation}
\Gamma(p;P) = - \int^\Lambda_k {\cal K}(p,k;P) S(k_+) \Gamma(k;P) S(k_-) \,,
\label{eq:BSE}
\end{equation}
where ${\cal K}(p,k;P)$ is the quark--antiquark kernel.

In RL truncation,
\begin{align}
{\cal K}(p,k;P) &= 4\pi \alpha_{\rm eff}(q^2) D^{\rm free}_{\mu\nu}(q) \nonumber \\
&
\times \left( \frac{\lambda^a}{2} \gamma_\mu \right) \otimes \left( \frac{\lambda^a}{2} \gamma_\nu \right) \,,
\end{align}
solutions exist only for discrete timelike values $P^2 = -M^2$, defining the meson mass $M$.

\subsection{Pseudoscalar and Vector Channels}

For pseudoscalar states ($0^{-+}$), the BS amplitude is expanded as
\begin{align}
\Gamma_{PS}(p;P) &= \gamma_5 \Big[ i E(p^2,p\!\cdot\!P) + \slashed{P} F(p^2,p\!\cdot\!P) \nonumber \\
&
+ \slashed{p} (p\cdot P) G(p^2,p\!\cdot\!P) \nonumber \\
&
+ \sigma_{\mu\nu} p_\mu P_\nu H(p^2,p\!\cdot\!P) \Big] \,.
\end{align}

For vector states ($1^{--}$),
\begin{equation}
\Gamma^\mu_V(p;P) = \sum_{i=1}^{8} F_i(p^2,p\!\cdot\!P) \, T^\mu_i(p;P) \,,
\end{equation}
where $T^\mu_i$ ($=T_\mu^i$) form a complete transverse basis:
\begin{subequations}
\begin{align}
T^1_\mu(q;P) &= \gamma^T_\mu \,,\\
T^2_\mu(q;P) &= \frac{6}{q^2\,\sqrt{5}} \Big[ q^T_\mu (\gamma^T\cdot q) - \case{1}{3} \gamma^T_\mu (q^T)^2 \Big] \,, \\
T^3_\mu(q;P) &= \frac{2}{q\,P} \Big[ q^T_\mu (\gamma\cdot P) \Big] \,,\\
T^4_\mu(q;P) &= \frac{i\,\sqrt{2}}{q\,P} \Big[ \gamma^T_\mu (\gamma\cdot P) (\gamma^T\cdot q) + q^T_\mu (\gamma\cdot P) \Big] \,, \\
T^5_\mu(q;P) &= \frac{2}{q}\;q^T_\mu \,,\\
T^6_\mu(q;P) &= \frac{i}{q\,\sqrt{2}} \Big[ \gamma_\mu^T (\gamma^T\cdot q) - (\gamma^T\cdot q) \gamma_\mu^T \Big] \,,\\ 
T^7_\mu(q;P) &= \frac{i\,\sqrt{3}}{q^2\,P\,\sqrt{5}} \left(1 - \cos^2{\theta}\right) \Big[ \gamma_\mu^T (\gamma\cdot P) \nonumber \\
&
- (\gamma\cdot P)\gamma_\mu^T \Big] - \frac{1}{\sqrt{2}}\;T^8_\mu(q;P) \,, \\
T^8_\mu(q;P) &= \frac{i\,2\sqrt{6}}{q^2\,P\,\sqrt{5}}\; q^T_\mu (\gamma^T\cdot q)(\gamma\cdot P) \,,
\end{align}
\end{subequations}
where $V^T$ is the component of $V$ transverse to $P$,
\begin{eqnarray}
V^T_\mu & = &  V_\mu - \frac{P_\mu\,(P\cdot V)}{P^2} \,.
\end{eqnarray}

\subsection{Decay Constants}

The leptonic decay constants are defined via
\begin{equation}
f_{PS} P_\mu = Z_2 \int^\Lambda_k {\rm Tr} \left[ \gamma_5 \gamma_\mu S(k_+) \Gamma_{PS}(k;P) S(k_-) \right] \,,
\end{equation}
for a pseudoscalar ($PS$) meson, and
\begin{equation}
f_V M_V = \frac{Z_2}{3} \int^\Lambda_k {\rm Tr} \left[ \gamma_\mu S(k_+) \Gamma^\mu_V(k;P) S(k_-) \right] \,.
\end{equation}
for a vector ($V$) meson.

These expressions are renormalisation-point independent when the truncation is symmetry preserving.


\section{RESULTS}
\label{sec:results}

\subsection{Running Mass and Renormalisation Scheme}
\label{subsec:Running}

The starting point of our analysis is the top-quark mass in the modified minimal subtraction ($\overline{\text{MS}}$) scheme, taken from Ref.~\cite{Beneke:2021lkq}:
\begin{equation}
m_t^{\overline{\text{MS}}}(m_t) = 163.643~\text{GeV}.
\end{equation}

Since continuum DSE approaches are typically formulated
in momentum-subtraction (MOM) schemes, it is convenient to relate the
$\overline{\text{MS}}$ mass to the corresponding MOM value.
At one loop in Landau gauge, the conversion reads~\cite{Celmaster:1979km, Chetyrkin:1999pq}:
\begin{equation}
m_t^{\text{MOM}} \simeq m_t^{\overline{\text{MS}}}
\left( 1 + \frac{4}{3}\frac{\alpha_s}{\pi} \right),
\quad (\text{1-loop, Landau gauge}) \,,
\end{equation}
with $\alpha_s$ the strong coupling constant. Note that, in the equation above, the masses and the coupling constant are evaluated at the same renormalisation scale.

The MOM renormalisation scheme is commonly employed in nonperturbative approaches to QCD, including lattice simulations and continuum functional methods such as DSEs. In this scheme, renormalisation constants are defined by requiring that propagators and vertex functions assume their tree-level forms at a fixed spacelike momentum $p^2=\mu^2$. Unlike the $\overline{\text{MS}}$ scheme, MOM renormalisation conditions depend explicitly on the gauge choice; however, they possess the advantage that the renormalisation scale $\mu$ is directly associated with the external momentum of the Green's functions, providing a more transparent physical interpretation.

The running of the $\overline{\text{MS}}$ mass is obtained from the renormalisation-group equation~\cite{Chetyrkin:1999pq}. Using the four-loop anomalous dimension, the scale dependence can be written as
\begin{align}
m_t^{\overline{\text{MS}}}(\mu^2) &= \hat m_t \left(\frac{\alpha_s}{\pi}\right)^{4/7} \Bigg[ 1 + 1.39796\left(\frac{\alpha_s}{\pi}\right) \nonumber\\
&
+ 1.79348\left(\frac{\alpha_s}{\pi}\right)^2 - 0.683433\left(\frac{\alpha_s}{\pi}\right)^3 \Bigg],
\end{align}
where $\hat m_t$ denotes the renormalisation-group invariant top-quark mass.

Table~\ref{tab:TopMassMOM} summarises the running of the top--quark mass and the conversion from the $\overline{\mathrm{MS}}$ scheme to the momentum subtraction (MOM) scheme for representative renormalisation scales in the range $\mu=400-800\,\text{GeV}$. As expected from the renormalisation-group evolution of QCD, both the strong coupling $\alpha_s(\mu)$ and the $\overline{\mathrm{MS}}$ mass decrease logarithmically as the renormalisation scale increases. The subsequent conversion to the MOM scheme, performed at one-loop order in Landau gauge, leads to slightly larger values for the running mass, reflecting the scheme dependence inherent to perturbative renormalisation prescriptions. Over the range of scales considered here the resulting MOM masses vary from approximately $160\,\text{GeV}$ to $152\,\text{GeV}$, corresponding to a moderate change of only a few percent. This relatively mild scale dependence indicates that the input quark mass used in the subsequent DSE-BSE calculations is stable against variations of the ultraviolet renormalisation point and provides a consistent starting point for the study of the $t\bar t$ bound-state system.

\begin{table}[!t]
\centering
\caption{\label{tab:TopMassMOM} Running of the top-quark mass and conversion from the $\overline{\mathrm{MS}}$ to the MOM scheme for representative renormalisation scales.}
\begin{ruledtabular}
\begin{tabular}{cccc}
$\mu$ (GeV) & $\alpha_s(\mu)$ & $m_t^{\overline{\mathrm{MS}}}(\mu)$ (GeV) & $m_t^{\mathrm{MOM}}(\mu)$ (GeV) \\
\hline
\tstrut
400 & 0.0976 & 153.308 & 159.660 \\
500 & 0.0953 & 150.999 & 157.104 \\
600 & 0.0934 & 149.184 & 155.098 \\
700 & 0.0919 & 147.696 & 153.455 \\
800 & 0.0906 & 146.440 & 152.071 \\
\end{tabular}
\end{ruledtabular}
\end{table}

\subsection{Charmonium and Bottomonium}
\label{subsec:Quarkonia}

Before addressing the hypothetical $t\bar{t}$ bound system, it is necessary to validate the present framework in sectors where experimental information is available. Heavy quarkonia, in particular charmonium ($c\bar c$) and bottomonium ($b\bar b$), provide ideal testing grounds for this purpose. These systems are sufficiently heavy that relativistic corrections remain moderate, while still probing the nonperturbative dynamics of QCD through the quark--gluon interaction kernel. Consequently, they have long served as benchmarks for continuum approaches based on DSE-BSE framework.

For systems composed solely of heavy quarks, the Qin--Chang effective interaction was adjusted in Ref.~\cite{Qin:2018dqp}, giving
\begin{equation}
\omega = 0.8\,\text{GeV} \,,\quad (\omega D)^3 = (0.6\,\text{GeV})^3 \,,
\end{equation}
where we remember that $D$ and $\omega$ control the infrared strength and width of the quark--gluon interaction. The number of active flavors is set to $N_f=5$ and thus $\gamma_m = 0.521739$ and $\Lambda_{\rm QCD}^{\overline{\text{MS}}}=0.21\,\text{GeV}$~\cite{ParticleDataGroup:2016lqr}. Note here the relation between MOM and $\overline{\text{MS}}$ renormalization schemes in the definition of $\Lambda_{\rm QCD}$~\cite{Boucaud:2008gn}:
\begin{equation}
\label{eq:MOMMS-LQCD}
\Lambda_{\text{QCD}}^{\text{MOM}} = \Lambda_{\text{QCD}}^{\overline{\text{MS}}} \cdot \exp\left(\frac{507-40N_f}{792-48N_f} \right) \,.
\end{equation}
which provides $\Lambda_{\text{QCD}}^{\text{MOM}}=0.36\,\text{GeV}$.

Using renormalised current-quark masses
\begin{align}
m_c(\mu=19\,\text{GeV}) &= 0.820\,\text{GeV} \,, \\
m_b(\mu=19\,\text{GeV}) &= 3.590\,\text{GeV} \,,
\end{align}
the Euclidean constituent quark masses are 
\begin{align}
M_c &= 1.322\,\text{GeV} \,, \\
M_b &= 4.215\,\text{GeV} \,,
\end{align}
and our theoretical framework reproduces the experimental masses and leptonic decay constants of the pseudoscalar and vector $S$-wave quarkonia, including the $\eta_c$, $J/\psi$, $\eta_b$, and $\Upsilon$ states (see Table~\ref{tab:Quarkonia1}).

\begin{table}[!t]
\caption{\label{tab:Quarkonia1} Computed values of masses and leptonic decay constants of the pseudoscalar and vector $S$-wave quarkonia, including $\eta_c$, $J/\psi$, $\eta_b$, and $\Upsilon$ states. Experimental values (Exp.) are taken from Ref.~\cite{ParticleDataGroup:2016lqr} -- $f_{\eta_c} = 0.238(12)$, $f_{J/\Psi} = 0.294(5)$, $f_{\Upsilon}=0.506(3)$; and lattice-QCD (Lat.) results are drawn from Refs.~\cite{Davies:2010ip, McNeile:2012qf, Donald:2012ga, Colquhoun:2014ica} -- $f_{\eta_c}=0.279(17)$, $f_{\eta_b}=0.472(4)$, $f_{J/\Psi}=0.286(4)$, $f_\Upsilon=0.459(22)$. All quantities in GeV.}
\begin{ruledtabular}
\begin{tabular}{l|rrr|rrr}
& \multicolumn{3}{c|}{Mass} & \multicolumn{3}{c}{Decay constant} \\[1ex]
& The. & Exp. & Lat. & The. & Exp. & Lat. \\
\hline
\tstrut
$\eta_c$ & 2.982 & 2.984 & $\cdots$ & 0.285 & 0.238 & 0.279 \\
$J/\psi$ & 3.123 & 3.097 & $\cdots$ & 0.301 & 0.294 & 0.286\\[2ex]
$\eta_b$   & 9.372 & 9.399 & $\cdots$ & 0.565 & $\cdots$ & 0.472 \\
$\Upsilon$ & 9.476 & 9.460 & $\cdots$ & 0.538 & 0.506 & 0.459 \\
\end{tabular}
\end{ruledtabular}
\end{table}

The successful description of these heavy-quark systems establishes the reliability of the interaction kernel and parameter set in the heavy-mass regime. In particular, the calculated masses of the lowest pseudoscalar and vector charmonium states agree with the experimental values at the level of a few percent, with $m_{\eta_c}$ and $m_{J/\psi}$ reproduced within approximately $1-3\%$. A similarly good agreement is obtained in the bottomonium sector, where the predicted masses of the $\eta_b$ and $\Upsilon$ states differ from the experimental values by less than about $1-2\%$. The corresponding leptonic decay constants are also consistent with phenomenological determinations and lattice-QCD results within the typical uncertainties expected in the rainbow--ladder truncation. Such quantitative agreement across both charmonium and bottomonium systems indicates that the effective interaction captures the dominant nonperturbative dynamics governing heavy quarkonia. This provides a controlled and well-tested starting point for extending the same framework to the extreme heavy-quark limit represented by the hypothetical $t\bar t$ bound system considered in the following subsection.

\subsection{Toponium}
\label{sec:toponium}

We now extend the same framework to the $t\bar t$ system. In order to preserve consistency with previous studies of heavy mesons within the Qin--Chang interaction, in our first analysis we consider the same model parameters, including $N_f=5$, corresponding to the standard situation in which the top quark does not contribute to the running of the QCD coupling below its mass threshold.

The dressed top-quark propagator is obtained from the gap equation using the running mass discussed in Sec.~\ref{subsec:Running}, and the BSE is solved in the pseudoscalar ($0^{-+}$) and vector ($1^{--}$) channels. Since the top mass lies far above the typical hadronic scale, it is useful to investigate the sensitivity of the bound-state solutions to the renormalisation point. We therefore compute the toponium masses and leptonic decay constants for a range of renormalisation scales $\mu = 400\text{--}800~\text{GeV}$, which are representative of the momentum region relevant for the heavy-quark dynamics entering the Bethe--Salpeter kernel. The resulting pseudoscalar and vector masses together with their leptonic decay constants are summarised in Table~\ref{tab:toponium1}.

\begin{table}[!t]
\centering
\caption{\label{tab:toponium1} Comparison of calculated toponium masses and leptonic decay constants with different renormalisation points. All
quantities listed in GeV.}
\begin{ruledtabular}
\begin{tabular}{cccccc}
$\mu$ & $M_t$ & $m_{t\bar t}^{PS}$ & $f_{t\bar t}^{PS}$ & $m_{t\bar t}^{V}$ & $f_{t\bar t}^{V}$  \\ 
\hline
\tstrut
400 & 163.803 & 344.537 & 7.080 & 344.708 & 6.674 \\ 
500 & 164.221 & 345.411 & 7.094 & 345.582 & 6.687 \\ 
600 & 164.083 & 345.154 & 7.120 & 345.326 & 6.714 \\
700 & 164.201 & 345.368 & 7.093 & 345.538 & 6.687 \\
800 & 164.485 & 345.992 & 7.134 & 346.164 & 6.728 \\
\end{tabular}
\end{ruledtabular}
\end{table}

Several qualitative features emerge from these results. First, the pseudoscalar and vector masses are nearly degenerate, with an average hyperfine splitting of $0.171\,\text{GeV}$, reflecting the expected heavy-quark spin symmetry in the extreme heavy-mass limit. Second, the leptonic decay constants are large, of order $f_{t\bar t}\sim 7~\text{GeV}$, consistent with the increasingly compact spatial structure of heavy quarkonia. Finally, the dependence on the renormalisation scale is mild, indicating that the predicted properties of the $t\bar t$ system are relatively stable within the range of scales considered.

Since the characteristic momentum scales relevant for the $t\bar t$ system lie well above the top mass, it is also instructive to investigate the case in which the top quark is treated as an active flavour in the renormalisation
group evolution.

The change in the number of active flavours modifies the ultraviolet behaviour of the interaction kernel through the quark-mass anomalous dimension and the QCD scale parameter. Following the relation between the MOM and $\overline{\text{MS}}$ schemes, Eq.~\eqref{eq:MOMMS-LQCD}, the value
\begin{equation}
\Lambda^{(6)}_{\text{MOM}} \simeq 0.151~\text{GeV}
\end{equation}
is obtained from the PDG value $\Lambda^{(6)}_{\overline{\text{MS}}}=0.089~\text{GeV}$.

When implementing $N_f=6$ within the Qin--Chang interaction, we retain the same infrared interaction parameters used in the previous subsection,
\begin{equation}
\omega = 0.8~\text{GeV}, \qquad
\omega D = (0.6~\text{GeV})^3 ,
\end{equation}
which ensures that the long-range structure of the kernel remains unchanged. However, the change in the ultraviolet running implies that the current heavy-quark masses must be slightly adjusted in order to reproduce the observed charmonium and bottomonium properties. Following this procedure we obtain
\begin{align}
m_c(\mu=19\,\text{GeV}) &= 0.873~\text{GeV} \,, \\
m_b(\mu=19\,\text{GeV}) &= 3.722~\text{GeV} \,,
\end{align}
which yield Euclidean constituent masses
\begin{align}
M_c &= 1.319\,\text{GeV} \,, \\
M_b &= 4.253\,\text{GeV} \,. 
\end{align}
With these parameters the calculated masses and leptonic decay constants of the $\eta_c$, $J/\psi$, $\eta_b$, and $\Upsilon$ states are shown in Table~\ref{tab:Quarkonia2} and, as one can notice, they remain in good agreement with experimental and lattice results, confirming that the interaction kernel continues to provide a reliable description of heavy quarkonia.

\begin{table}[!t]
\caption{\label{tab:Quarkonia2} With $N_f=6$ and thus fine-tuning the Qin--Chang effective interaction, computed values of masses and leptonic decay constants of the pseudoscalar and vector $S$-wave quarkonia, including $\eta_c$, $J/\psi$, $\eta_b$, and $\Upsilon$ states. Experimental values (Exp.) are taken from Ref.~\cite{ParticleDataGroup:2016lqr} -- $f_{\eta_c} = 0.238(12)$, $f_{J/\Psi} = 0.294(5)$, $f_{\Upsilon}=0.506(3)$; and lattice-QCD (Lat.) results are drawn from Refs.~\cite{Davies:2010ip, McNeile:2012qf, Donald:2012ga, Colquhoun:2014ica} -- $f_{\eta_c}=0.279(17)$, $f_{\eta_b}=0.472(4)$, $f_{J/\Psi}=0.286(4)$, $f_\Upsilon=0.459(22)$. All quantities in GeV.}
\begin{ruledtabular}
\begin{tabular}{l|rrr|rrr}
& \multicolumn{3}{c|}{Mass} & \multicolumn{3}{c}{Decay constant} \\[1ex]
& The. & Exp. & Lat. & The. & Exp. & Lat. \\
\hline
\tstrut
$\eta_c$ & 2.978 & 2.984 & $\cdots$ & 0.260 & 0.238 & 0.279 \\
$J/\psi$ & 3.087 & 3.097 & $\cdots$ & 0.273 & 0.294 & 0.286 \\[2ex]
$\eta_b$   & 9.399 & 9.399 & $\cdots$ & 0.488 & $\cdots$ & 0.472 \\
$\Upsilon$ & 9.463 & 9.460 & $\cdots$ & 0.466 & 0.506 & 0.459 \\
\end{tabular}
\end{ruledtabular}
\end{table}

Using this updated parameter set we then solve the BSE for the $t\bar t$ system over the same range of renormalisation scales. The resulting pseudoscalar and vector toponium masses and decay constants are presented in Table~\ref{tab:toponium2}. The pseudoscalar and vector masses are slightly smaller in this case, with an average hyperfine splitting of $0.135\,\text{GeV}$. The predicted leptonic decay constants remain large, $f_{t\bar t}\sim 6$--$6.5~\text{GeV}$, although they are slightly smaller than the corresponding $N_f=5$ values due to the modified ultraviolet running of the interaction. Finally, the dependence on the renormalisation scale with $N_f=6$ is even milder than the $N_f=5$ case.

\begin{table}[!t]
\centering
\caption{\label{tab:toponium2} With $N_f=6$ and thus fine-tuning the Qin--Chang effective interaction, comparison of calculated toponium masses and leptonic decay constants with different renormalisation points. All quantities listed in GeV.}
\begin{ruledtabular}
\begin{tabular}{cccccc}
$\mu$&$M_t$& $m_{t\Bar{t}}^{PS}$  & $f_{t\Bar{t}}^{PS}$ & $m_{t\Bar{t}}^{V}$ & $f_{t\Bar{t}}^{V}$  \\ 
\hline
\tstrut
$400$ & 163.634 & 343.432 & 6.495 & 343.567 & 6.123 \\ 
$500$ & 163.968 & 344.119 & 6.504 & 344.255 & 6.132 \\ 
$600$ & 163.781 & 343.743 & 6.491 & 343.878 & 6.152 \\
$700$ & 163.857 & 343.879 & 6.493 & 344.014 & 6.122 \\
$800$ & 164.106 & 344.415 & 6.510 & 344.551 & 6.137 \\
\end{tabular}
\end{ruledtabular}
\end{table}

Quantitatively, the inclusion of the top quark as an active flavour leads to a modest reduction in the predicted decay constants, $\sim\!(8-10)\%$, and a slight modification of the bound-state masses, $\sim\!1\,\text{GeV}$. This behaviour can be traced to the weaker effective interaction arising from the modified renormalisation-group running when $N_f=6$, which reduces the binding strength in the Bethe--Salpeter kernel. Nevertheless, the overall structure of the spectrum remains remarkably stable under this change, indicating that our results are largely insensitive to the precise choice of active flavour number in the ultraviolet evolution, although quantitative differences at the level of a few percent may arise in the predicted masses and decay constants.


\section{SUMMARY AND OUTLOOK}
\label{sec:summary}

In this work we have investigated the properties of the lowest-lying pseudoscalar ($0^{-+}$) and vector ($1^{--}$) $t\bar t$ systems within a fully Poincar\'e--covariant Dyson--Schwinger--Bethe--Salpeter (DSE--BSE) framework. The analysis employs the rainbow--ladder truncation together with the Qin--Chang effective interaction, an approach that has previously demonstrated quantitative success in describing meson properties from the light-quark sector up to heavy quarkonia.

As a necessary validation step, we first confirmed that the interaction kernel and parameter set reproduce the masses and leptonic decay constants of the charmonium and bottomonium ground states with good accuracy. This provides confidence that the same framework can be consistently extrapolated to the extreme heavy-quark regime represented by the top quark.

Using the running top-quark mass defined in the modified minimal subtraction $\overline{\mathrm{MS}}$ scheme and converted to the momentum-subtraction (MOM) scheme appropriate for continuum functional methods, we solved the quark DSE and the homogeneous BSE for the $t\bar t$ system. The analysis was performed for renormalisation scales in the range $\mu = 400$--$800~\mathrm{GeV}$ and for two different choices of active flavour number, $N_f=5$ and $N_f=6$, allowing us to quantify the sensitivity of the results to ultraviolet renormalisation-group evolution.

Our calculations predict pseudoscalar and vector toponium masses in the range
\begin{equation}
m_{t\bar t} \sim 344-346\,\text{GeV} \,,
\end{equation}
with a very small hyperfine splitting of order
\begin{equation}
\Delta m_{\text{HF}} \sim 0.14-0.17\,\text{GeV} \,,
\end{equation}
reflecting the expected emergence of heavy--quark spin symmetry in the extreme heavy-mass limit. The corresponding leptonic decay constants are found to be large,
\begin{equation}
f_{t\bar t} \sim 6-7\,\text{GeV} \,,
\end{equation}
consistent with the increasingly compact spatial structure of heavy quarkonia as the quark mass grows.

A comparison between the $N_f=5$ and $N_f=6$ scenarios reveals that increasing the number of active flavours slightly weakens the effective interaction through the modified renormalisation-group running. This leads to a modest reduction in both the predicted masses and decay constants, at the level of a few percent, but the overall structure of the spectrum remains remarkably stable. Likewise, the dependence on the renormalisation point in the range considered is mild, indicating that the predicted properties of the $t\bar t$ system are robust against variations of the ultraviolet scale.

Although the physical top quark decays weakly with a lifetime shorter than the typical hadronisation time, the present results demonstrate that the underlying QCD dynamics still supports a strongly correlated quark--antiquark configuration in the pseudoscalar and vector channels. In this sense, the $t\bar t$ system may be interpreted as representing the ultimate heavy-quark limit of quarkonium binding within the Standard Model. Such correlations could manifest themselves experimentally through threshold enhancements or quasi-bound structures in top-pair production processes, as suggested by recent analyses of LHC data.

Several extensions of the present work would further refine this picture. A first important improvement would consist in incorporating the finite electroweak width of the top quark, which would require treating the quark propagator with a complex pole structure and solving the Bethe--Salpeter equation in the presence of unstable constituents. In addition, corrections beyond the rainbow--ladder truncation, such as dressed quark--gluon vertices and crossed-ladder contributions, may become relevant for achieving higher quantitative precision. Finally, the present study focuses on the ground-state pseudoscalar and vector channels; extending the analysis to excited states and other quantum numbers could provide a more complete description of the possible spectrum of correlated $t\bar t$ configurations.


\begin{acknowledgments}
Work supported by:
Natural Science Foundation of Jiangsu Province (grant no. BK20220122); 
National Natural Science Foundation of China (grant no. 12233002);
and Ministerio Espa\~nol de Ciencia e Innovaci\'on, grant no. PID2022-140440NB-C22.
\end{acknowledgments}


\bibliography{print_Toponium-CSM}

\end{document}